\documentstyle[sprocl,epsfig,wrapfig,amsmath]{article}

\begin{document}

\pagestyle{empty}

\begin{flushleft}
\Large
{SAGA-HE-140-98
\hfill Oct. 20, 1998}  \\
\end{flushleft}
 
\vspace{1.6cm}
 
\begin{center}
 
\LARGE{{\bf Transversity distributions}} \\
\vspace{0.2cm}

\LARGE{{\bf  and spin asymmetries}} \\

\vspace{1.1cm}
 
\LARGE
{S. Hino, M. Hirai, S. Kumano, and M. Miyama $^*$} \\
 
\vspace{0.3cm}
  
\LARGE
{Department of Physics}         \\
 
\LARGE
{Saga University}      \\
 
\LARGE
{Saga 840-8502, Japan} \\

\vspace{1.0cm}
 
\LARGE
{Talk given at the 13th International Symposium} \\

\vspace{0.1cm}

{on High Energy Spin Physics} \\

\vspace{0.1cm}

{Protvino, Russia, Sept. 8 -- 12, 1998} \\

\vspace{0.05cm}

{(talk on Sept. 8, 1998) }  \\
 
\end{center}
 
\vspace{0.7cm}

\vfill
 
\noindent
{\rule{6.0cm}{0.1mm}} \\
 
\vspace{-0.3cm}
\normalsize
\noindent
{* Email: 97sm16@edu.cc.saga-u.ac.jp, 
98td25@edu.cc.saga-u.ac.jp,} \\

\vspace{-0.44cm}
\noindent
{\ \ \ kumanos@cc.saga-u.ac.jp,
96td25@edu.cc.saga-u.ac.jp.} \\

\vspace{-0.44cm}
\noindent
{\ \ \ Information on their research is available 
 at http://www2.cc.saga-u.ac.jp}  \\

\vspace{-0.44cm}
\noindent
{\ \ \ /saga-u/riko/physics/quantum1/structure.html.} \\

\vspace{+0.1cm}
\hfill
{\large to be published in proceedings by the World Scientific}

\vfill\eject
\setcounter{page}{1}
\pagestyle{plain}

%%%%%%%%%%%%%%%%%%%%%%%%%%%%%%%%%%%%%%%%%%%%%%%%%%%%%%%%%%%%%%%%%%%%%%%%%%%%%%%
%%%%%%%%%%%%%%%%%%%%%%%%%%%%%%%%%%%%%%%%%%%%%%%%%%%%%%%%%%%%%%%%%%%%%%%%%%%%%%%

\title{TRANSVERSITY DISTRIBUTIONS \\ AND SPIN ASYMMETRIES}

\author{S. HINO, M. HIRAI, S. KUMANO, M. MIYAMA}

\address{Department of Physics, Saga University \\ 
         Saga 840-8502, Japan}

\maketitle\abstracts{
We discuss transversity distributions and transverse spin asymmetries.
First, $Q^2$ evolution results are shown for transversity and
longitudinally polarized distributions.
Second, the antiquark flavor asymmetry 
$\Delta_{_T} \bar u/\Delta_{_T} \bar d$
is discussed in two different descriptions, a meson-cloud model and
an exclusion model. Both calculations produce a significant
$\Delta_T \bar d$ excess over $\Delta_{_T} \bar u$. Third, we explain
general formalism of the polarized proton-deuteron Drell-Yan
processes. This study is partly intended to find the asymmetry
$\Delta_{_T} \bar u/\Delta_{_T} \bar d$ by the p-d asymmetry.
However, the existence of tensor structure in the deuteron produces
a number of new structure functions. There exist 108 structure
functions and 22 ones even after integrating the cross section
over the virtual-photon transverse momentum $\vec Q_T$.
}

%%%%%%%%%%%%%%%%%%%%%%%%%%%%%%%%%%%%%%%%%%%%%%%%%%%%%%%%%%%%%%%%%%%%%%%%%%%%%%%
%%%%%%%%%%%%%%%%%%%%%%%%%%%%%%%%%%%%%%%%%%%%%%%%%%%%%%%%%%%%%%%%%%%%%%%%%%%%%%%
\section{Introduction}
\label{sec:intro}

Using many experimental data on $g_1$, we have a rough idea on
the longitudinally polarized parton distributions.
However, the transversity distributions $\Delta_{_T} q$
have not been measured at all because they do not contribute
to the cross section of inclusive lepton-proton scattering. 
They are expected to be measured by
the Drell-Yan process at RHIC. It is, therefore, important to
predict the features of $\Delta_{_T} q$ as many as we can
before the RHIC experiment.
There are three major reasons for studying the transversity distributions.
First, our knowledge of proton spin should be tested by another
observable. Second, the transversity $Q^2$ evolution is very different
from the longitudinal one. It is a good test of perturbative QCD.
Third, nonrelativistic quark models predict that transversity
distributions are equal to the corresponding longitudinally polarized
ones, so that the difference could shed light on the relativistic aspect
of hadron structure.

We first discuss the $Q^2$ evolution of a transversity
distribution.\cite{h1} Then, possible mechanisms are explained
for producing the light antiquark flavor asymmetry 
$\Delta_{_T} \bar u/\Delta_{_T} \bar d$.\cite{sk,km}
Obtained results are used for calculating the transverse
double spin asymmetry at a RHIC energy.\cite{km} 
The ratio $\Delta_{_T} \bar u/\Delta_{_T} \bar d$ could be extracted
from the pp and pd Drell-Yan data. However, because there had been
no available formalism for the polarized pd Drell-Yan, we explain
various spin asymmetries for finding the structure functions of
a spin-one hadron.\cite{pd}

%%%%%%%%%%%%%%%%%%%%%%%%%%%%%%%%%%%%%%%%%%%%%%%%%%%%%%%%%%%%%%%%%%%%%%%%%%%%%%%
%%%%%%%%%%%%%%%%%%%%%%%%%%%%%%%%%%%%%%%%%%%%%%%%%%%%%%%%%%%%%%%%%%%%%%%%%%%%%%%
\section{$Q^2$ evolution and light antiquark flavor asymmetry}
\label{sec:trans}

%%%%%%%%%%%%%%%%%%%%%%%%%%%%%%%% figure %%%%%%%%%%%%%%%%%%%%%%%%%%%%%%%%%%%%%%
\begin{wrapfigure}{r}{0.46\textwidth}
   \begin{center}
   \epsfig{file=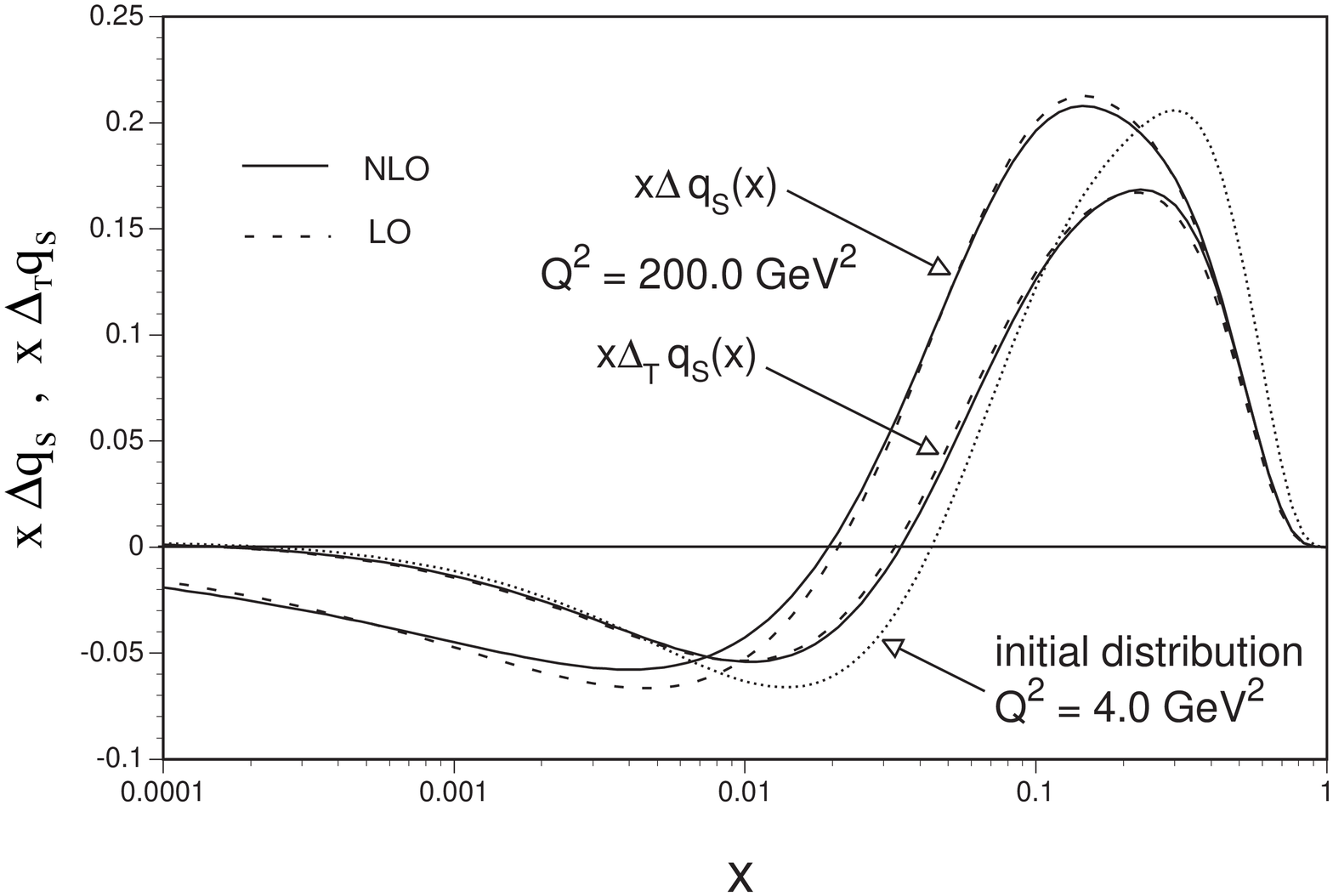,width=5.0cm}
   \end{center}
   \vspace{-0.5cm}
       \caption{\footnotesize $Q^2$ evolution.}
       \label{fig:sing}
\end{wrapfigure}
%%%%%%%%%%%%%%%%%%%%%%%%%%%%%%%% figure %%%%%%%%%%%%%%%%%%%%%%%%%%%%%%%%%%%%%%
In order to test perturbative QCD, we have to rely on
the $Q^2$ dependence of structure functions. 
Fortunately, the next-to-leading-order $Q^2$ evolution has been
completed in the recent years not only for the longitudinally
polarized distributions but also for the transversity ones.
It was revealed that the transversity $Q^2$ evolution is very different
from the longitudinal one because $\Delta_{_T} q$ does not
couple to the gluon distribution.\cite{h1}
We show the singlet evolution results in Fig. \ref{fig:sing}
by using our program in Ref. 1.\footnote{
   Our $Q^2$ evolution program could be obtained upon email request.
   \\ \ \ \!  
See http://www2.cc.saga-u.ac.jp/saga-u/riko/physics/quantum1/program.html.}
The evolved distributions are significantly different even though
the initial ones are identical. The difference should be an important
test of perturbative QCD in spin physics.

%%%%%%%%%%%%%%%%%%%%%%%%%%%%%%%% figure %%%%%%%%%%%%%%%%%%%%%%%%%%%%%%%%%%%%%%
\begin{wrapfigure}{r}{0.46\textwidth}
   \begin{center}
   \epsfig{file=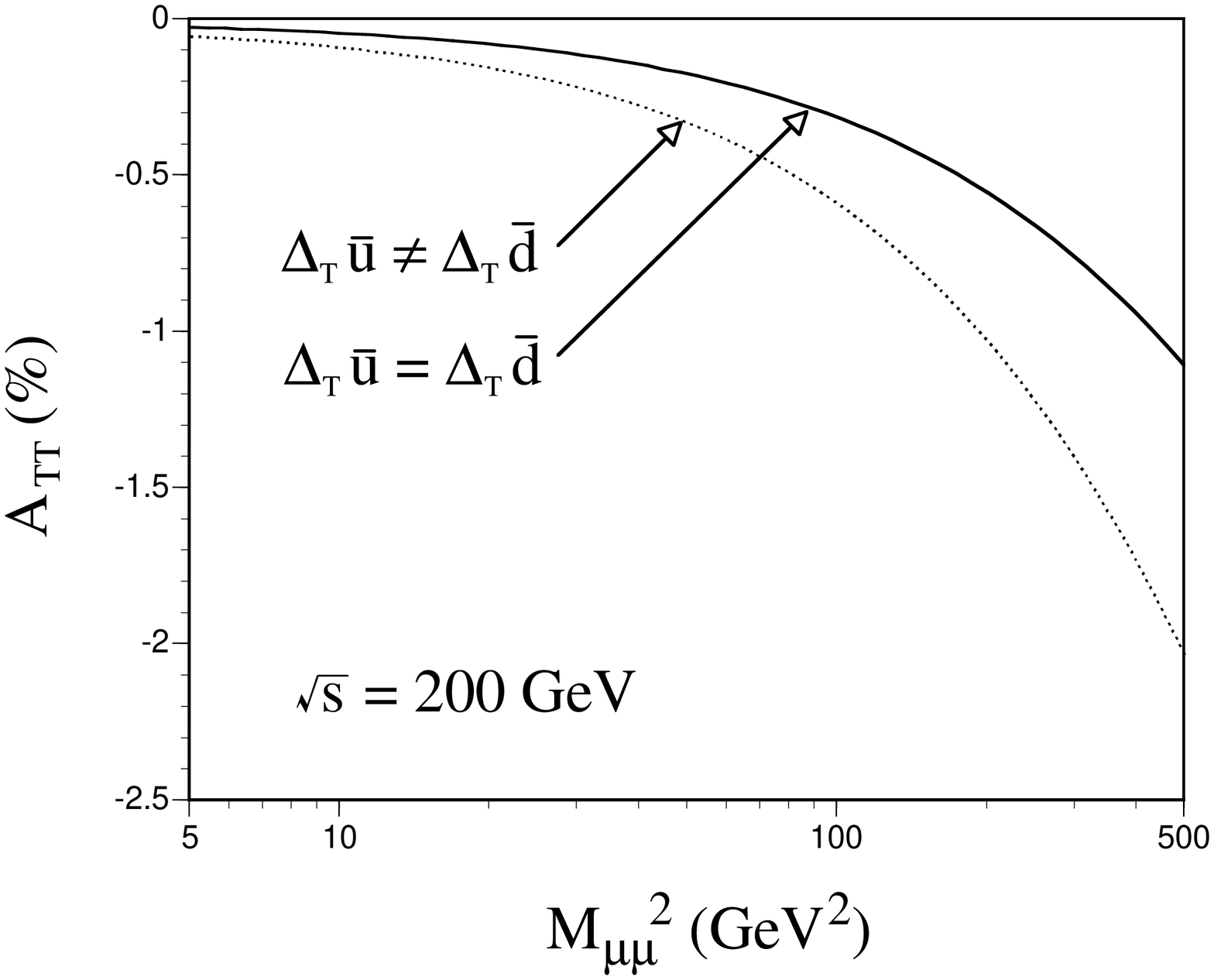,width=5.0cm}
   \end{center}
   \vspace{-0.6cm}
       \caption{\footnotesize Transverse spin asymmetry.}
       \label{fig:att}
\end{wrapfigure}
%%%%%%%%%%%%%%%%%%%%%%%%%%%%%%%% figure %%%%%%%%%%%%%%%%%%%%%%%%%%%%%%%%%%%%%%
There are studies on the possible antiquark flavor asymmetry
$\Delta \bar u/\Delta \bar d$ or
$\Delta_{_T} \bar u/\Delta_{_T} \bar d$.\cite{sk,km}
We have looked at two different descriptions, a meson-cloud model and
an exclusion model.\cite{km} The $\rho$-meson clouds could contribute to
the polarized flavor asymmetries, for example, because $\rho^+$ contains
the $\bar d$-valence quark. The exclusion principle is another
approach. It produces the spin asymmetry, for example, because
it is more difficult to create $u^\uparrow$ than to create $u^\downarrow$ 
due to the $u^\uparrow$ excess over $u^\downarrow$ in the SU(6)
quark model. It is interesting to find that both models predict
$\Delta_{_T} \bar d$ excess over $\Delta_{_T} \bar u$ although
their magnitude could be different. We show the transverse double
asymmetries calculated by using the flavor symmetric parton distributions
and by the exclusion model in Fig. \ref{fig:att}.\cite{km}
At this stage, the numerical analysis of the meson-cloud model is
still preliminary so that its results are not shown here.
The two curves are significantly different; however,
the measurement of $A_{TT}$ is not enough to find the ratio
$\Delta_{_T} \bar u/\Delta_{_T} \bar d$. Although the longitudinal
ratio $\Delta \bar u/\Delta \bar d$ should be found by the W production
processes, the transversity ratio cannot be measured by the W. 
As the alternative, we propose to use the polarized
proton-deuteron Drell-Yan process in combination with
the polarized pp process.

\vfill\eject
%%%%%%%%%%%%%%%%%%%%%%%%%%%%%%%%%%%%%%%%%%%%%%%%%%%%%%%%%%%%%%%%%%%%%%%%%%%%%%%
%%%%%%%%%%%%%%%%%%%%%%%%%%%%%%%%%%%%%%%%%%%%%%%%%%%%%%%%%%%%%%%%%%%%%%%%%%%%%%%
\section{Polarized proton-deuteron Drell-Yan processes}
\label{sec:pd}

The major purpose for studying polarized proton-deuteron (pd) Drell-Yan
processes is to investigate new spin-dependent structure functions for
spin-one hadrons. However, the flavor asymmetry
$\Delta_T \bar u/\Delta_T \bar d$ could be investigated as a byproduct.
General formalism for the polarized pd reactions was completed
recently.\cite{pd}
Imposing Hermiticity, parity conservation, and time-reversal invariance, 
we have found that 108 structure functions exist in the pd Drell-Yan
processes. The number reduces to 22 after integrating over
the virtual-photon transverse momentum $\vec Q_T$ or after taking
the limit $Q_T\rightarrow 0$. The following spin asymmetries could
be investigated in the pd reactions:
\begin{alignat}{8}
& < \! \sigma \! >, \ \ & & 
A_{LL}, \ \             & &
A_{TT}, \ \             & &
A_{LT}, \ \             & &
A_{TL}, \ \             & &
A_{UT}, \ \             & &
A_{TU}, \ \             & &
        \ \        \nonumber \\
& A_{UQ_0}, \ \         & &     
A_{TQ_0}, \ \           & &
A_{UQ_1}, \ \           & &
A_{LQ_1}, \ \           & &
A_{TQ_1}, \ \           & &
A_{UQ_2}, \ \           & &
A_{LQ_2}, \ \           & &
A_{TQ_2},
\nonumber
\end{alignat}
where the subscripts $U$, $L$, and $T$ indicate unpolarized,
longitudinally polarized, and transversely polarized states.
The superscripts $Q_0$, $Q_1$, and $Q_2$ are quadrupole polarizations,
and they are associated with the spherical harmonics $Y_{20}$, $Y_{21}$,
and $Y_{22}$. Because the deuteron is a spin-one hadron, there are
structure functions which cannot be studied in the proton-proton
reactions. There are 11 new structure functions in addition to the 11 ones
in the Drell-Yan processes of spin-1/2 hadrons, 
and they can be measured by the quadrupole polarization asymmetries.
The deuteron reaction may be realized in the RHIC-Spin project
and other future ones. Because the essential formalism is now completed,
it is also possible to study the asymmetry 
$\Delta_T \bar u/\Delta_T \bar d$ 
by measuring the transverse asymmetry $A_{TT}$ and then
by taking the difference between the polarized pp and pd cross sections.
However, we should be careful about the way to measure 
the transverse asymmetry in the sense that the tensor structure
functions may contribute.\cite{pd}

\vspace{0.3cm}

M.H. and M.M. were supported by the JSPS Research Fellowships
for Young Scientists.
S.K. was partly supported by the Grant-in-Aid from the Japanese Ministry
of Education, Science, and Culture under \#10640277. 

\vspace{-0.1cm}
%%%%%%%%%%%%%%%%%%%%%%%%%%%%%%%%%%%%%%%%%%%%%%%%%%%%%%%%%%%%%%%%%%%%%%%%%%%%%%%
%%%%%%%%%%%%%%%%%%%%%%%%%%%%%%%%%%%%%%%%%%%%%%%%%%%%%%%%%%%%%%%%%%%%%%%%%%%%%%%

\end{document}